\documentclass[runningheads]{llncs}
\usepackage[T1]{fontenc}
\usepackage{algorithm}
\usepackage{algorithmic}
\usepackage{cite}
\usepackage{amsmath,amssymb,amsfonts}
\usepackage{graphicx}
\usepackage{textcomp}
\usepackage{comment}
\usepackage{booktabs}
\usepackage{caption}
\usepackage{balance}
\usepackage{subfig}  

\usepackage{graphicx}
\begin{document}
\title{A Granularity Characterization of Task Scheduling Effectiveness}

\author{Sana Taghipour Anvari \and David Kaeli}

\authorrunning{S. Taghipour Anvari and D. Kaeli}

\institute{Northeastern University, Boston, MA, USA \\
\email{taghipouranvari.s@northeastern.edu}}

\maketitle

\begin{abstract}
Task-based runtime systems provide flexible load balancing and portability for parallel scientific applications, but their strong scaling is highly sensitive to task granularity. As parallelism increases, scheduling overhead may transition from negligible to dominant, leading to rapid drops in performance for some algorithms, while remaining negligible for others. Although such effects are widely observed empirically, there is a general lack of understanding how algorithmic structure impacts whether dynamic scheduling is always beneficial. In this work, we introduce a granularity characterization framework that directly links scheduling overhead growth to task-graph dependency topology. We show that dependency structure, rather than problem size alone, governs how overhead scales with parallelism. Based on this observation, we characterize execution behavior using a simple granularity measure that indicates when scheduling overhead can be amortized by parallel computation and when scheduling overhead dominates performance. Through experimental evaluation on representative parallel workloads with diverse dependency patterns, we demonstrate that the proposed characterization explains both gradual and abrupt strong-scaling breakdowns observed in practice. We further show that overhead models derived from dependency topology accurately predict strong-scaling limits and enable a practical runtime decision rule for selecting dynamic or static execution without requiring exhaustive strong-scaling studies or extensive offline tuning.
\end{abstract}

\section{Introduction}
\label{sec:introduction}

Task-based runtimes are widely used in parallel scientific computing. By expressing computations as a directed acyclic graph of tasks with data dependencies, these systems enable flexible load balancing, overlap of computation and communication, and portable parallelism across diverse hardware. As a result, task-based runtime systems have been adopted in application domains such as computational fluid dynamics~\cite{carpaye2017cfd} and molecular dynamics~\cite{acun2017namd}. However, practitioners face a persistent question when deploying task-based systems at scale: \emph{when does dynamic scheduling help, and when does it hurt?} Every dynamic scheduler imposes nonzero per-task overhead for dependency tracking, queue management, and work-stealing decisions. When tasks are large, this overhead can be amortized by the amount of work that will benefit. When tasks are smaller, the overhead can hide any benefits obtained from scheduling. We commonly encounter applications that transition between these two environments. Even though dynamic scheduling has been adopted widely in practice~\cite{starpu,parsec,legion,charmpp,hpx}, prior work has not actively explored the reasons why applications experience varying degrees of overhead, impacting scaling behavior.

The problem is especially acute under strong scaling. A distributed application that benefits from task scheduling at moderate rank counts may experience degradation as the domain is further decomposed. As per-task work shrinks and the number of scheduling decisions grows, overhead that was once negligible can dominate execution time. Prior experimental studies of task-based runtime systems report scheduling overhead fractions ranging from under 2\% to over 90\%, depending on the workload, runtime, and execution configuration~\cite{daggerfft,zhang2023quantifying,hpx}. Different workloads can reach this threshold at vastly different scales. Understanding this disparity requires examining the connection between algorithm structure and scheduler behavior.

In this paper, we show that scheduling overhead can be estimated by inspecting the task-graph dependency topology. This insight provides a unified explanation for the widely varying overhead behaviors reported across different workloads and scales. We make three contributions:

\begin{enumerate}
    \item Algorithm-based overhead estimation.
    We demonstrate that scheduling overhead scaling depends not only on problem size but also on the dependency topology of the task graph, and that different dependency structures lead to qualitatively different overhead growth rates under strong scaling.

    \item Structure-driven overhead model.
    We introduce a simple scheduling overhead model whose functional form is determined by the dependency topology of the task graph, defined by the pattern of inter-task dependencies across ranks.

    \item Unified characterization framework. We introduce a workload-independent metric for characterizing scheduling effectiveness that identifies distinct execution ranges and enables a practical decision rule for selecting between dynamic and static execution.

    \item Experimental validation. We evaluate the proposed framework across representative workloads spanning global, local, and independent dependency classes, demonstrating that the overhead model accurately predicts strong-scaling limits and that all configurations collapse onto a single workload-independent relationship.
\end{enumerate}

The remainder of this paper is organized as follows. Section~\ref{sec:related} reviews prior work on task granularity and scheduling overhead in task-based runtime systems. Section~\ref{sec:background} introduces the runtime model and dependency topology concepts that motivate our analysis. Section~\ref{sec:characterizing} presents the granularity number, the overhead model, and the resulting regime classification. Section~\ref{sec:evaluation} evaluates the framework across representative workloads and validates the proposed characterization under strong scaling. Section~\ref{sec:summary} concludes with practical implications and guidance for using dynamic scheduling at scale.

\section{Related Work}
\label{sec:related}

Our work addresses the question of \emph{when} dynamic task scheduling provides benefit over static execution in distributed scientific computing. We position our contributions relative to three bodies of research: i) task granularity characterization, ii) scheduler overhead modeling, and iii) adaptive scheduling strategies.

\paragraph{Task Granularity and Scheduling Overhead Characterization} Task granularity and scheduler overhead have been studied extensively in the context of task-based runtime systems. Task Bench~\cite{taskbench} introduced the Minimum Effective Task Granularity (METG) metric across a broad range of programming systems, demonstrating that tasks smaller than approximately 100~$\mu$s lead to substantial efficiency loss. Subsequent studies have reported similar thresholds in specific runtimes. Grubel \textit{et al.}~\cite{hpx-tasksize} characterized scheduling overheads in HPX as a function of task granularity and showed that thread management overhead dominates execution time for fine-grained tasks, while Zhang \textit{et al.}~\cite{zhang2023quantifying} quantified scheduling overheads in Charm++~\cite{charmpp} and HPX~\cite{hpx} using Task Bench. HyperQueue~\cite{hyperqueue} further demonstrated that lightweight task management can sustain per-task overheads of approximately 0.1~ms while scaling to millions of tasks. Scheduler latency has also been examined at coarser time scales. Reuther \textit{et al.}~\cite{reuther2018scalable} modeled scheduler behavior for HPC and Big Data systems, showing that resource utilization drops sharply when scheduling latency becomes comparable to computation time. Related work in uncertainty quantification workflows~\cite{uq-workflow-scheduling} reported up to three orders of magnitude reduction in scheduling overhead by replacing batch schedulers with fine-grained task schedulers, highlighting the importance of scheduler choice for latency-sensitive workloads.

\paragraph{Task-Based Runtime Systems}
Task-based runtimes exhibit distinct overhead characteristics that define their effective operating regimes. StarPU~\cite{starpu} demonstrates that dynamic scheduling guided by performance estimates improves execution time on heterogeneous systems by reducing load imbalance and selecting efficient device mappings, without explicitly modeling task granularity. PaRSEC~\cite{parsec,parsec-ecp} reduces scheduling overhead through Parameterized Task Graphs that avoid explicit graph materialization and enables efficient execution of fine-grained tasks. TTG~\cite{ttg} further targets low-overhead dataflow execution by avoiding synchronization bottlenecks on shared runtime data structures, allowing very small tasks to execute efficiently. Legion and Regent~\cite{legion,regent} are two frameworks designed to mitigate fine-grained overhead through region-based dependence analysis and compiler-driven task fusion, achieving substantial speedups on irregular applications. nOS-V~\cite{nosv} explicitly identifies multiple operating regimes, including a crossover between peak-performance and overhead-dominated execution. 

\paragraph{Load Balancing and Work Stealing}
Work stealing algorithms face the same granularity–overhead tradeoff that we study. PackStealLB~\cite{packsteallb} reduces overhead by grouping tasks into packs, improving performance by up to 41\% on molecular dynamics benchmarks by amortizing communication costs that would dominate fine-grained individual migrations. Chung et al.~\cite{proactive-offloading,proactive-offloading-journal} show that reactive work stealing can be ineffective when communication delays are high, and propose proactive offloading guided by load prediction, achieving 1.5–3.4$\times$ speedups. Scalable work stealing~\cite{scalable-work-stealing} demonstrates high efficiency up to 8,192 processors, but notes that performance depends critically on the computation-to-communication ratio of stolen tasks.

\paragraph{Adaptive Scheduling Strategies}
Several systems propose adaptive mechanisms to mitigate scheduling and task management overhead, addressing challenges related to workload variability. HeraSched~\cite{herasched} applies hierarchical reinforcement learning to job selection and allocation in HPC systems and demonstrates substantial reductions in job waiting time under overloaded conditions. Agon~\cite{agon2021} employs a neural network classifier to select among multiple schedulers, achieves up to 99.1\% of oracle performance, illustrating that the most effective scheduling strategy depends on workload characteristics. ATG~\cite{atg} adaptively controls the creation of tasks in work-stealing schedulers and reduces overhead by switching between help-first and serialized execution, resulting in performance improvements of up to 19\%. StarPU's hierarchical tasks~\cite{starpu-hierarchical} enable dynamic granularity adaptation to better match heterogeneous resources. The taskiter directive~\cite{taskiter} amortizes task creation overhead in iterative applications by reusing task graphs across iterations, which results in average speedups of 3.7$\times$ for fine-grained workloads. Finally, DAG clustering~\cite{dag-clustering} increases task granularity through aggregation, reduces overhead, and achieves up to 7$\times$ speedup. 

\paragraph{Comparison With Our Work}
Prior work quantifies minimum effective task granularity and scheduling overheads across runtimes~\cite{taskbench,hpx-tasksize,zhang2023quantifying,hyperqueue} and proposes mechanisms to mitigate fine-grained costs through dataflow optimizations, packing, or clustering~\cite{parsec,ttg,legion,regent,packsteallb,dag-clustering}. Adaptive approaches further select or adjust scheduling strategies at runtime~\cite{atg,starpu-hierarchical,herasched,agon2021}, but none provides a predictive rule for \emph{when} dynamic task scheduling should be preferred over static execution for distributed scientific workloads under varying scale and overhead conditions. We address this gap by linking overhead scaling rates to task-graph dependency topology, characterizing the resulting regime transition, and using it to choose between dynamic and static execution in our distributed benchmarks.

\section{Background and Motivation}
\label{sec:background}

Task-based runtime systems are increasingly adopted for distributed scientific computing, yet practitioners consistently report a recurring challenge: at moderate scales, dynamic scheduling improves performance through load balancing and latency hiding, but at higher scales, the same scheduling mechanisms can become the dominant cost. This tension between scheduling benefit and scheduling overhead motivates the central question of this work: \emph{when does dynamic scheduling help, and when does it hurt?}

A per-phase runtime model~\cite{daggerfft} captures the interaction between useful computation, communication, and scheduling overhead within a single computation stage of task-based execution. The model is expressed as
\begin{equation}
T_{\text{phase}} \approx \max(T_{\text{comp}}, T_{\text{comm}}) + (1 - \rho)\, k\, \tau_s
\label{eq:daggerfft}
\end{equation}
where $T_{\text{comp}}$ denotes useful computation time, $T_{\text{comm}}$ denotes communication time (e.g., MPI transfers or data redistributions), $k$ is the number of scheduled tasks per worker, $\tau_s$ is the per-task scheduling cost including dependency tracking, queue management, and dispatch, and $\rho \in [0,1]$ represents the fraction of scheduling overhead that can be hidden by parallel execution. The term $(1 - \rho) k \tau_s$ therefore represents the scheduler overhead that remains exposed on the critical path.

In practice, communication and computation are not strictly separated into distinct phases. In pipelined task-based execution, data redistribution between stages overlaps with task execution across ranks. No global barrier separates redistribution from subsequent computation; ranks enter and exit phases asynchronously, with each rank's progress gated only by the arrival of its required input data. As a result, communication costs are not a separable term but are embedded in the pipeline stall time that each rank experiences while waiting for inter-rank data dependencies to resolve. We therefore model the total non-kernel cost per phase as a single overhead term:
\begin{equation}
T_{\text{phase}} \approx T_{\text{kernel}} + T_{\text{overhead}}
\label{eq:phase_reduced}
\end{equation}
where $T_{\text{overhead}}$ captures the aggregate dependency cost, including pipeline stalls from inter-rank data dependencies, redistribution management, and scheduler dispatch (dependency tracking, queue management, task placement). The functional form of $T_{\text{overhead}}$ is determined by the dependency topology of the algorithm, which governs both the number of inter-rank data dependencies per phase and the resulting network contention. Previous work~\cite{daggerfft, taskbench, hpx-tasksize, nosv} reports that under strong scaling, as task granularity decreases, overhead can transition abruptly from negligible to dominant, overwhelming useful computation and producing a sharp performance cliff rather than gradual degradation. This behavior suggests a threshold below which per-task dependency costs can no longer be amortized effectively. We refer to this phenomenon as the \emph{overhead cliff}.

\subsection{Dependency Topology in Distributed Algorithms}
Distributed algorithms exhibit fundamentally different dependency structures depending on how tasks relate across ranks. We characterize these structures using a dependency neighborhood $\mathcal{N}(t,i)$, which specifies the set of tasks from timestep $t-1$ that task $(t,i)$ depends on~\cite{taskbench}. Table~\ref{tab:dependence} summarizes the three dependency classes.

\begin{table}[t]
\centering
\caption{Dependency neighborhood relations and their complexity for representative algorithm classes, where $P$ is the number of ranks and $i$ is the task index.\cite{taskbench}}
\label{tab:dependence}
\begin{tabular*}{1.0\linewidth}{@{\extracolsep{\fill}}lccc}
\toprule
\textbf{Pattern} & \textbf{Dependency Neighborhood} & \textbf{Deps/Task} & \textbf{Total Edges} \\
\midrule
Global (FFT) & $\mathcal{N}(t,i) := \{j \mid 0 \leq j < P\}$ & $O(P)$ & $O(P^2)$ \\
Local (Stencil) & $\mathcal{N}(t,i) := \{i, i-1, i+1\}$ & $O(1)$ & $O(P)$ \\
Local (Sweep) & $\mathcal{N}(t,i) := \{i, i-1\}$ & $O(1)$ & $O(P)$ \\
Independent (GEMM) & $\mathcal{N}(t,i) := \emptyset$ & $O(0)$ & $O(1)$ \\
\bottomrule
\end{tabular*}
\end{table}

\textbf{Global dependency} algorithms induce all-to-all dependency structures. The classic example is distributed FFT~\cite{cooleyfft,p3dfft, czechowski2012fft}, where each computation phase consists of $P$ local 1D transforms whose inputs depend on data from all other ranks. This creates $O(P)$ dependencies per task and $O(P^2)$ total edges in the dependency graph.

\textbf{Local dependency} algorithms require only nearest-neighbor communication. Stencil computations with halo exchange~\cite{datta2008stencil} exemplify this pattern: each subdomain requires boundary data only from adjacent subdomains in the processor topology. Sweep computations exhibit a similar bounded dependency structure, where each task depends only on a fixed number of neighboring tasks along a traversal direction. These patterns create $O(1)$ dependencies per task (at most six neighbors in 3D for stencil), yielding $O(P)$ total edges.

\textbf{No dependency} algorithms consist of independent tasks after the initial data distribution. Block matrix multiplication with a 2D block-cyclic distribution~\cite{vandegeijn1997summa} exhibits this pattern: after initial data placement, each rank performs local computation with no runtime synchronization, which results in $O(1)$ total dependency overhead, regardless of scale.

\begin{figure}[t]
    \centering
    \includegraphics[width=\linewidth]{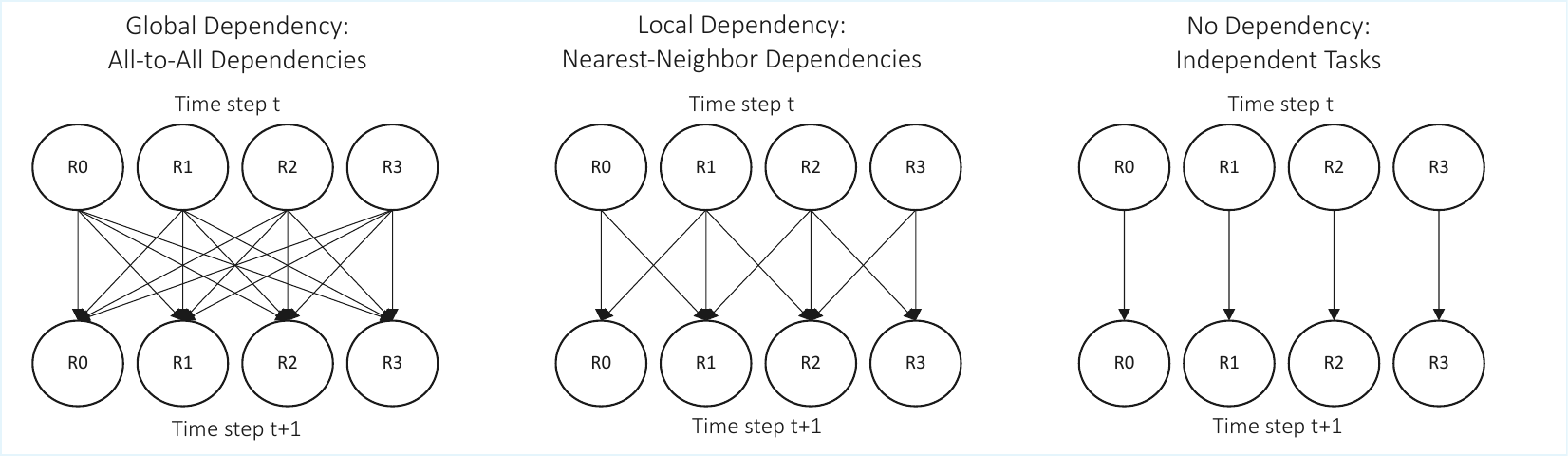}
    \caption{Illustrative dependency topologies at $P=4$ ranks.
    Edges represent logical inter-rank dependencies that constrain task execution order between computation phases. Global dependency (left) induces all-to-all dependencies, local dependency (center) induces nearest-neighbor dependencies, and independent workloads (right) induce no runtime dependencies.}
    \label{fig:coordination_topology}
\end{figure}

Figure~\ref{fig:coordination_topology} illustrates these dependency patterns for $P=4$ ranks and highlights how global, local, and independent workloads induce fundamentally different dependency structures between computation phases. These three topology classes span the spectrum of dependency structures in scientific computing and suggest that scheduler overhead scales with the number of inter-rank dependency edges rather than task count alone. Section~\ref{sec:characterizing} formalizes this relationship and derives the resulting overhead scaling behavior for each class.

\section{Characterizing Scheduler Overhead}
\label{sec:characterizing}

The scheduling overhead behavior under strong scaling is determined by the dependency structure of the task graph rather than task count or problem size alone. This distinction explains why some workloads experience abrupt performance collapse while others degrade gradually or remain stable.

Starting from the reduced phase-level model in Eq.\ref{eq:phase_reduced}, we isolate the component of execution time attributable to task scheduling:
\begin{equation}
T_{\text{overhead}} = (1 - \rho)\, k\, \tau_s,
\label{eq:toverhead}
\end{equation}
and define the granularity number as the ratio of useful computation to scheduling overhead:

\begin{equation}
G \equiv \frac{T_{\text{kernel}}}{T_{\text{overhead}}}
= \frac{T_{\text{kernel}}}{(1 - \rho)\, k\, \tau_s}.
\label{eq:granularity}
\end{equation}

Large values of $G$ indicate that kernel execution dominates and scheduling overhead is negligible. Values near unity correspond to situations where overhead and computation are comparable. When $G < 1$, scheduling overhead exceeds useful computation and the scheduler becomes the dominant performance bottleneck. The overhead fraction, which is the ratio of scheduling time to kernel time, follows directly as $\Omega = 1/G$.

\subsection{Dependency Topology Determines Overhead Scaling}
Scheduling overhead scales with the number of inter-task dependency edges in the task graph: $T_{\text{overhead}}(P) \propto |E(P)|$, where $E(P)$ denotes the set of directed edges representing inter-task dependencies at scale $P$. Because different algorithms induce qualitatively different dependency structures, they exhibit distinct overhead scaling behaviors:

\textbf{Global dependency.} Algorithms that require all-to-all dependencies induce dense task graphs where each task depends on many remote tasks. The number of dependency edges grows superlinearly: $E_{\text{global}}(P) = \Theta(P^2)$.
Scheduling overhead increases rapidly under strong scaling, eventually overwhelming useful computation.

\textbf{Local dependency.} Algorithms with nearest-neighbor or bounded-radius dependencies induce sparse dependency graphs. Each task depends on a constant number of neighbors:
$E_{\text{local}}(P) = \Theta(P)$. Scheduling overhead grows linearly with scale, producing gradual degradation rather than abrupt collapse.

\textbf{No runtime dependency.} Algorithms whose tasks are independent during the computation phase induce constant dependency complexity: $E_{\text{independent}}(P) = \Theta(1)$. Scheduling overhead remains largely insensitive to parallel scale.

\subsection{Granularity Decay Under Strong Scaling}

Under strong scaling with fixed global problem size, we assume ideal scaling of a fully parallelizable computational kernel with negligible intrinsic serial fraction. The kernel execution time therefore decreases inversely with the number of ranks,
\begin{equation}
T_{\text{kernel}}(P) = \frac{A}{P},
\end{equation}
where $A$ is a constant dependent on the architecture and implementation.

Scheduling overhead scales with the dependency burden imposed by the task-graph dependency structure. Because the scheduler must track and resolve inter-task dependency edges, we model overhead growth as proportional to the total number of such edges,
\begin{equation}
T_{\text{overhead}}(P) \propto |E(P)|,
\end{equation}
where $|E(P)|$ is determined by the dependency topology.

Using the definition of the granularity number in Eq.\ref{eq:granularity}, we obtain:
\begin{equation}
G(P) \propto \frac{P^{-1}}{|E(P)|}.
\end{equation}

Substituting the asymptotic growth of $E(P)$ for each topology class yields:

\begin{align}
E_{\text{global}}(P) &= \Theta(P^2) 
\;\Rightarrow\;
G_{\text{global}}(P) \propto P^{-3} \quad \text{(rapid collapse)}, \\
E_{\text{local}}(P) &= \Theta(P) 
\;\Rightarrow\;
G_{\text{local}}(P) \propto P^{-2} \quad \text{(gradual degradation)}, \\
E_{\text{independent}}(P) &= \Theta(1) 
\;\Rightarrow\;
G_{\text{independent}}(P) \propto P^{-1} \quad \text{(stable scaling)}.
\end{align}

The granularity number therefore decays at rates determined entirely by coordination topology. Dense global coordination induces cubic decay under strong scaling, bounded local coordination induces quadratic decay, and independent workloads induce linear decay.

A natural transition scale $P^\star$ arises when useful computation and scheduling overhead become comparable, $T_{\text{kernel}}(P^\star) = T_{\text{overhead}}(P^\star)$, or equivalently $G(P^\star)=1$. The location of $P^\star$ depends directly on how rapidly $E(P)$ grows: global-dependency algorithms reach this transition quickly, while local and independent algorithms delay or avoid it within practical scaling ranges.

\section{Evaluation}
\label{sec:evaluation}
We validate the proposed granularity characterization using a set of representative workloads with distinct dependency structures. Table~\ref{tab:experimental_setup} summarizes the experimental platform and workload configurations used in this study. For each configuration, we measure the kernel execution time $T_{\text{kernel}}$ and the scheduling overhead $T_{\text{overhead}}$, and compute $G$ (Eq.\ref{eq:granularity}) to identify the overhead range.

\begin{table}[t]
\centering
\caption{Experimental platform and workloads.}
\label{tab:experimental_setup}
\begin{tabular}{p{3.5cm} p{8.5cm}}
\toprule
\textbf{Category} & \textbf{Configuration} \\
\midrule
Hardware &
CPU cluster with Intel Xeon Gold 6240R processors (48 cores per node, 2.4\,GHz base clock, 36\,MB L3 cache), connected via InfiniBand HDR. \\
Software &
Julia\cite{julia} v1.11.7; FFTW\cite{fftw, fftw.jl} v3.3.10 for local transforms; Open MPI\cite{mpi.jl} v4.1.6 with MPI.jl v0.20.22; Dagger.jl\cite{dagger} v0.18.14 for task scheduling. \\

\midrule
FFT Workload &
3D complex-to-complex FFT. Problem sizes $N \in \{384^3, 512^3, 768^3\}$, scales 4--256 ranks. Each computation phase spawns $P$ tasks with all-to-all dependencies, inducing $\Theta(P^2)$ dependency edges. \\
Stencil Workload &
Five-point stencil on a 2D grid with 1D row decomposition. Problem sizes $N \in \{7525^2, 11560^2, 21285^2\}$ across 4--256 ranks. Each iteration spawns $P$ tasks with nearest-neighbor dependencies, inducing $\Theta(P)$ dependency edges. \\
Sweep Workload &
Discrete ordinates transport sweep on a 2D grid with 1D row decomposition. Problem sizes $N \in \{7525^2, 11560^2, 21285^2\}$ across 4--256 ranks. Each iteration spawns $P$ tasks with unidirectional left-to-right dependency, inducing $\Theta(P)$ dependency edges. \\
GEMM Workload &
Blocked matrix multiplication with 1D row distribution. Problem sizes $N \in \{7525^2, 11560^2, 21285^2\}$, scales 4--256 ranks. After initial data distribution, tasks execute independently, inducing $\Theta(1)$ dependency edges. \\
\bottomrule
\end{tabular}
\end{table}

\subsection{Contrasting Failure Modes}
\begin{figure}[t]
    \centering
    \includegraphics[width=\columnwidth]{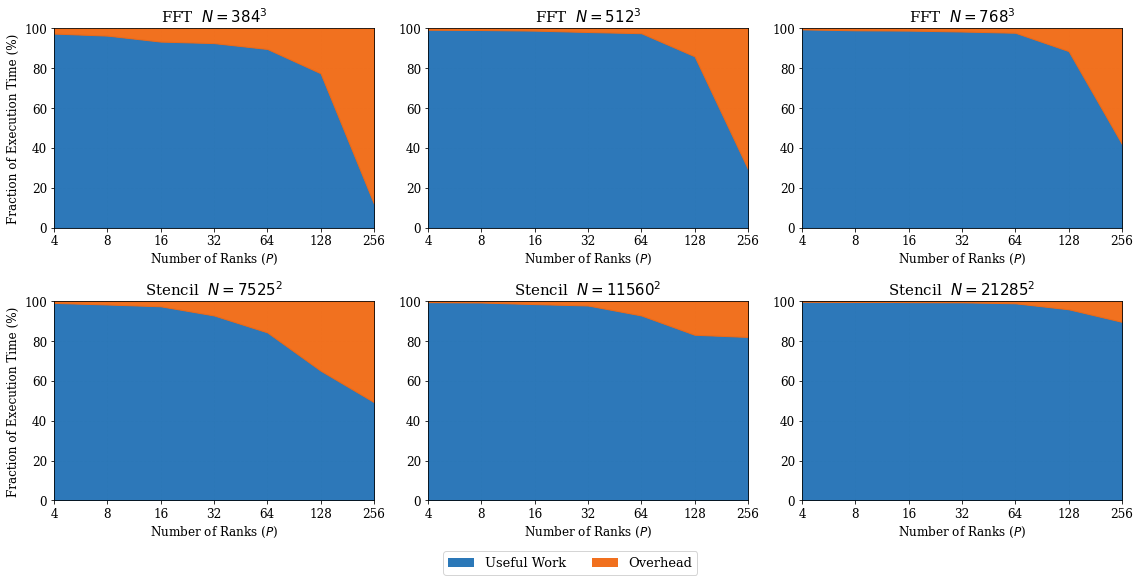}
    \caption{Contrasting failure modes under strong scaling with dynamic scheduling. FFT (top row) exhibits an overhead cliff where scheduling overhead remains negligible across moderate scales but rises abruptly at high rank counts, consuming a majority of execution time at 256 ranks. Stencil (bottom row) exhibits gradual degradation where overhead grows steadily with scale and only approaches parity with useful work at the highest ranks.}
    \label{fig:failure_modes}
\end{figure}

Figure~\ref{fig:failure_modes} illustrates how dependency topology impacts scheduling overhead under strong scaling. We plot the fraction of execution time spent on useful computation versus scheduling overhead for dynamically scheduled FFT and stencil workloads across all three problem sizes.

Across all FFT problem sizes, overhead remains below 10\% through 64 ranks, then increases sharply at higher scales. At 256 ranks, scheduling overhead dominates execution time, consuming 50--80\% depending on problem size. In contrast, stencil workloads exhibit smooth, monotonic overhead growth from under 1\% at small scales to tens of percent at the highest ranks. Despite comparable problem sizes and task counts, these two workloads exhibit fundamentally different overhead trajectories, suggesting that dependency structure rather than scale alone determines how overhead grows.

\subsection{Granularity--Overhead Relationship}

\begin{figure}[t]
    \centering
    \includegraphics[width=0.8\columnwidth]{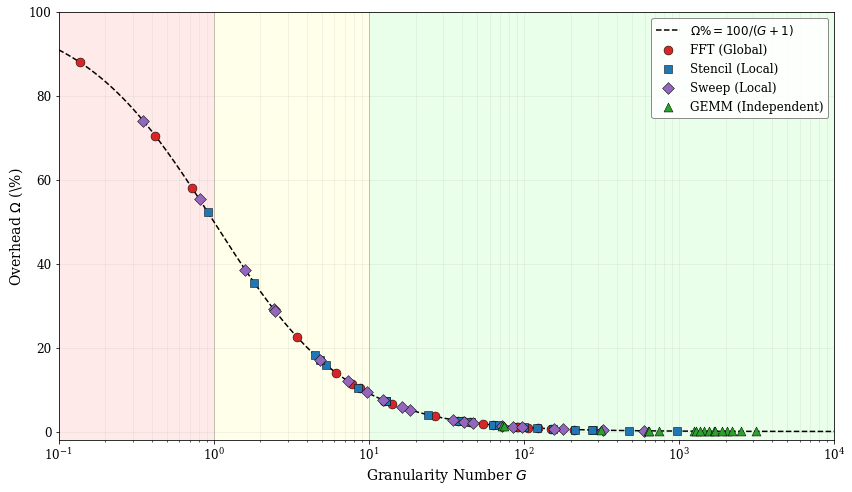}
    \caption{Overhead fraction versus granularity number $G$ for all measured configurations. Each point represents one experimental configuration. Background shading indicates execution regimes: detrimental (red, $G < 1$), marginal (yellow, $1 \leq G < 10$), and beneficial (green, $G \geq 10$). While all configurations obey the same relationship, their trajectories under strong scaling differ: FFT (circles) traverses all three regimes, stencil and sweep (squares and diamonds) reach the marginal regime at high rank counts, and GEMM triangles) remains entirely in the beneficial regime.}
    \label{fig:universal_scaling}
\end{figure}

Figure~\ref{fig:universal_scaling} plots the fraction of execution time spent in overhead, $\Omega\%$, as a function of the granularity number $G$~\eqref{eq:granularity} for all measured configurations across FFT, stencil, sweep, and GEMM workloads. Since $T_{\text{phase}} = T_{\text{kernel}} + T_{\text{overhead}}$, the overhead fraction follows directly as
\begin{align}
\Omega\% &=
\frac{T_{\text{overhead}}}{T_{\text{kernel}} + T_{\text{overhead}}} \times 100 
= \frac{100}{G + 1}.
\label{eq:omega}
\end{align}
All measurements collapse onto this relationship, confirming that $G$ provides a single, workload-independent characterization of scheduling effectiveness regardless of algorithmic structure, problem size, or scale.

The granularity number induces a natural transition boundary at $G = 1$, where overhead equals useful computation ($\Omega = 50\%$). We further adopt $G = 10$ as a practical threshold for negligible overhead ($\Omega \approx 9\%$), consistent with the convention that parallel efficiency losses below 10\% are acceptable in practice~\cite{taskbench,hpx-tasksize}.
We classify execution into three ranges based on these boundaries. In the \textbf{beneficial} range ($G > 10$), overhead remains below ${\sim}9\%$ and dynamic scheduling provides net performance benefit. In the \textbf{marginal} range ($1 < G \leq 10$), overhead becomes comparable to computation and the benefits of dynamic scheduling diminish. In the \textbf{detrimental} range ($G \leq 1$), overhead exceeds useful computation and static execution is preferable.

While all configurations obey the same $\Omega\%$--$G$ relationship, their \emph{trajectories} under strong scaling differ fundamentally. Independent workloads (GEMM) consistently exhibit large granularity values ($G \gtrsim 70$), remaining firmly in the beneficial range across all tested scales. Stencil and sweep workloads transition gradually into the marginal range only at high rank counts, reflecting the slower granularity decay ($G \propto P^{-2}$) imposed by bounded-degree inter-rank dependencies. FFT workloads span all three ranges, with the rapid decay ($G \propto P^{-3}$) driven by dense inter-rank dependencies causing an abrupt transition to the detrimental range at the largest scales. These distinct trajectories demonstrate that the dependency topology governs \emph{where} a workload operates on the universal curve and \emph{how quickly} it moves toward the detrimental range as parallelism increases.

\subsection{Model Calibration and Validation}
\label{sec:calibration}

We evaluate the proposed direct scheduling-overhead model by calibrating it against measured runtime data and validating its ability to identify strong-scaling limits across workloads and input sizes. Rather than inferring overhead indirectly through efficiency metrics, we model scheduling cost explicitly as a function of rank count~$P$, that reflects the dependency structure imposed by each workload and is observed directly in the runtime behavior.

\begin{figure}[tbp]
    \centering
    \includegraphics[width=1\linewidth]{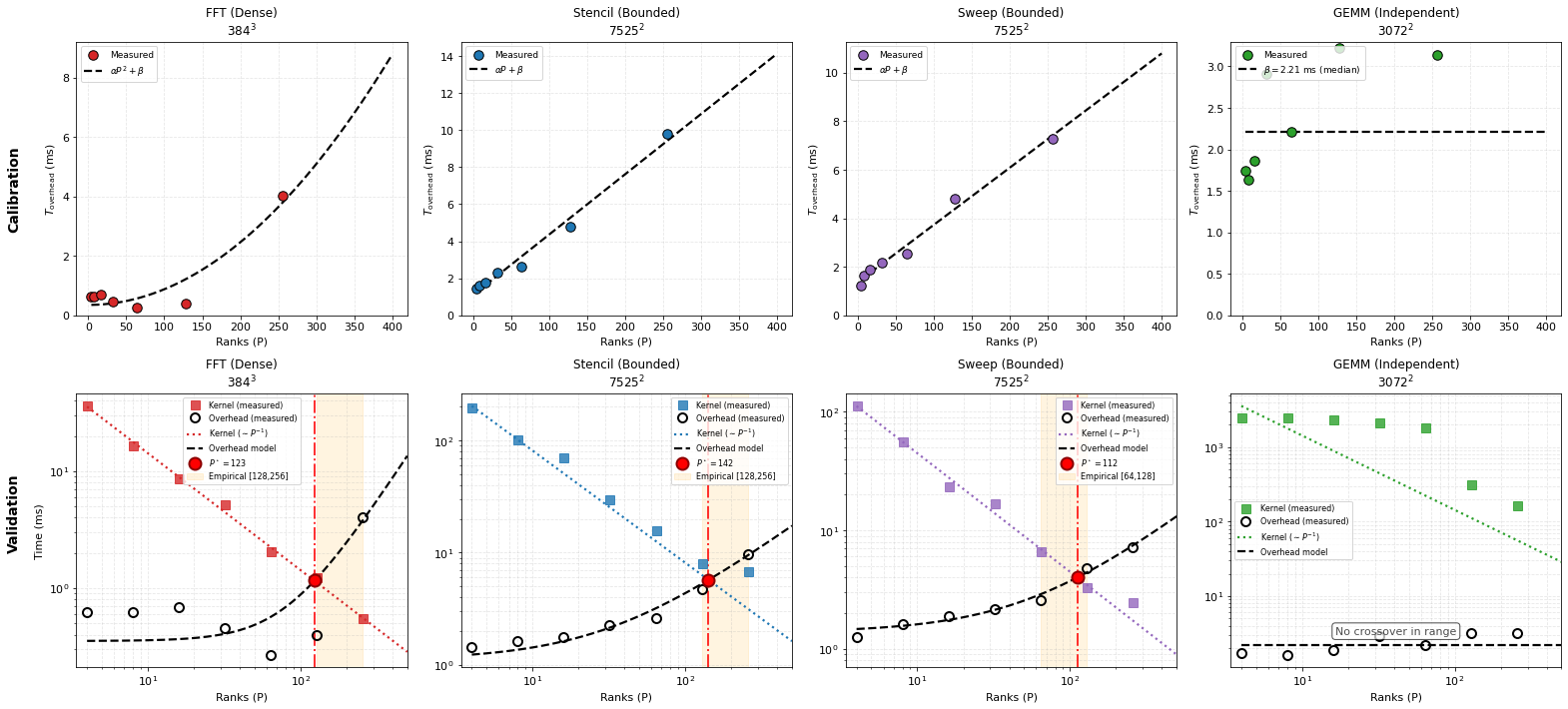}\par\vspace{0.3em}
    \includegraphics[width=1\linewidth]{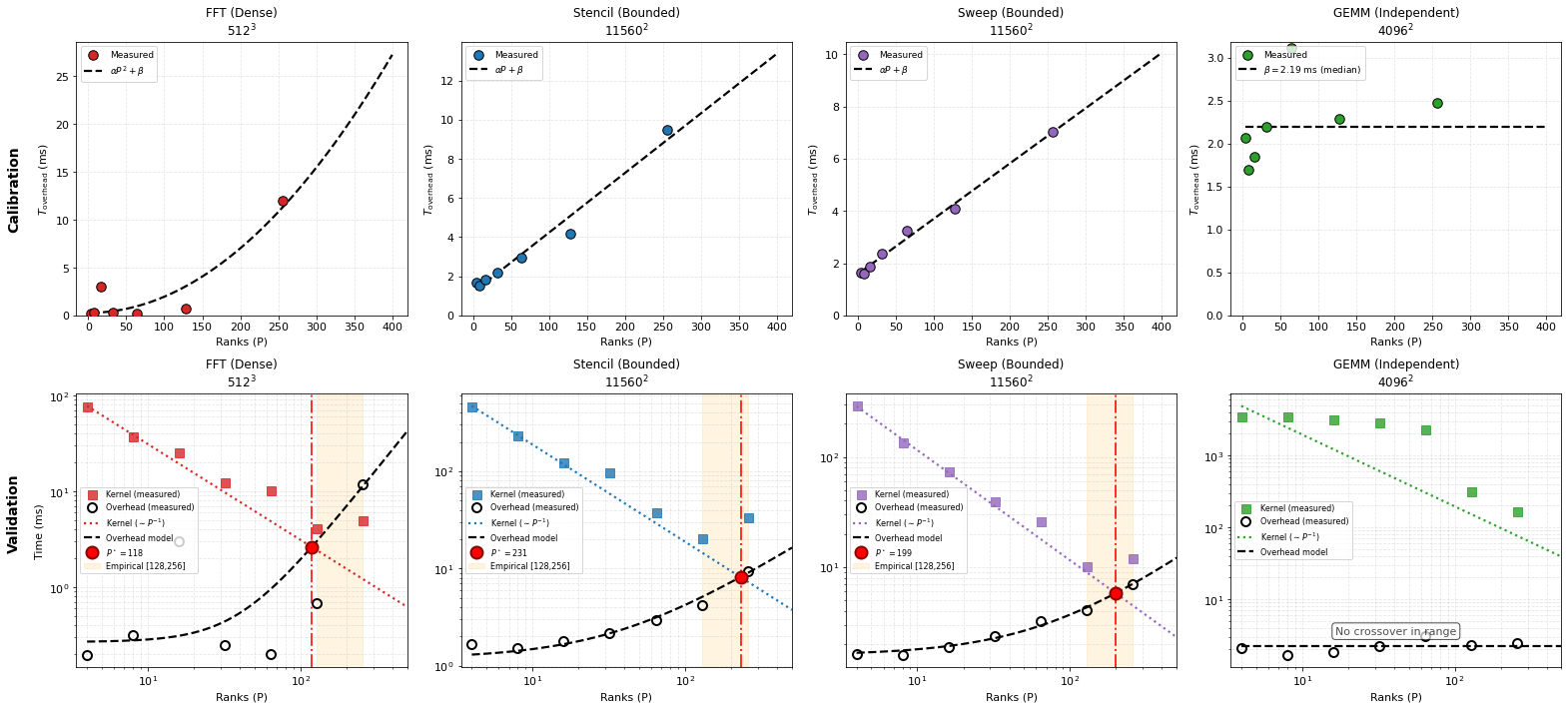}\par\vspace{0.3em}
    \includegraphics[width=1\linewidth]{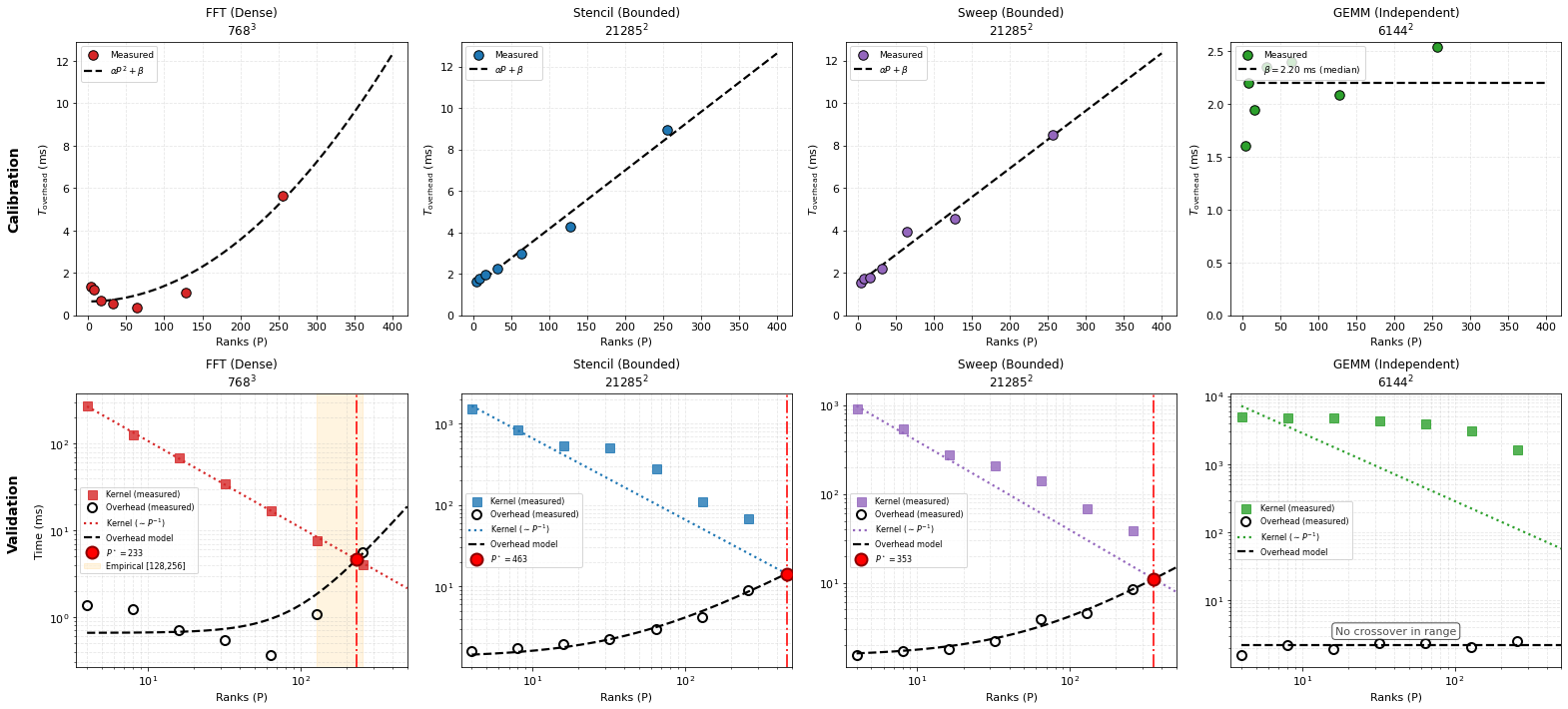}
    \caption{Calibration and validation of the direct scheduling-overhead model across three input sizes. Top panels fit the overhead model; bottom panels overlay it against kernel strong scaling. The red dot marks the predicted crossover $P^\star$. Shaded regions indicate the empirical interval between the last non-detrimental and first detrimental ($G < 1$) measured configuration. Where observed, $P^\star$ falls within or adjacent to this interval.
    }
    \label{fig:direct_overhead_calibration}
\end{figure}

Figure~\ref{fig:direct_overhead_calibration} summarizes calibration and validation results for three problem sizes. Each row corresponds to a different input size, while columns correspond to workloads with distinct dependency patterns: FFT (global dependency), stencil (local dependency), sweep (local dependency), and GEMM (independent tasks). For each input size, the top panels calibrate the scheduling-overhead model, and the bottom panels validate the model under strong scaling.

\paragraph{Calibration.}
We model the total scheduling overhead $T_{\mathrm{overhead}}(P)$ directly as a function of the rank count~$P$, using functional forms dictated by the dependency structure of each workload:
\begin{align}
\text{global dependency:} \quad
& T_{\mathrm{overhead}}(P) = \alpha P^2 + \beta, \label{eq:fft_overhead} \\
\text{local dependency:} \quad
& T_{\mathrm{overhead}}(P) = \alpha P + \beta, \label{eq:stencil_overhead} \\
\text{independent:} \quad
& T_{\mathrm{overhead}}(P) = \beta. \label{eq:gemm_overhead}
\end{align}

Model parameters are obtained via least-squares fitting to the measured scheduling-overhead times.
For FFT, calibration is restricted to the pre-collapse scaling region ($P \leq P^\star$) to ensure the model captures overhead growth before it dominates execution. For GEMM and sweep, the constant term~$\beta$ is estimated using the median measured overhead to ensure robustness to noise.

\paragraph{Validation and strong-scaling limit.}
To validate the calibrated overhead models, we compare them against ideal kernel strong scaling,
\begin{equation}
T_{\mathrm{kernel}}(P) = \frac{A}{P},
\label{eq:kernel_model}
\end{equation}
where $A$ is obtained by fitting measured kernel execution times.

We define the strong-scaling limit $P^\star$ as the point at which scheduling overhead equals useful computation:
\begin{equation}
T_{\mathrm{overhead}}(P^\star) = T_{\mathrm{kernel}}(P^\star).
\label{eq:cliff_condition}
\end{equation}

As shown in Figure~\ref{fig:direct_overhead_calibration}, FFT exhibits a quadratic growth in scheduling overhead that intersects the $P^{-1}$ kernel scaling curve at a finite $P^\star$, predicting the onset of an overhead-dominated regime whose location depends on input size.
This crossover consistently aligns with the empirically observed transition from marginal to detrimental strong-scaling behavior.
In contrast, the kernel--overhead intersection for stencil workloads occurs at substantially larger rank counts than for FFT, and scheduling overhead does not induce an abrupt performance collapse within the measured scaling range. Notably, stencil's absolute overhead in milliseconds can exceed FFT's at comparable scales, yet its overhead \emph{fraction} remains much lower (Fig.~\ref{fig:failure_modes}) because stencil kernels are significantly larger. This illustrates that the overhead regime is determined by the ratio $G = T_{\text{kernel}}/T_{\text{overhead}}$, not by absolute overhead alone.
For GEMM, no kernel--overhead intersection is observed within the measured scaling range, consistent with its independent task structure and near-constant scheduling overhead.

\subsection{Predicting Strong-Scaling Limits Across Dependency Classes}

To evaluate the generality of the proposed overhead model beyond the workloads used for calibration in Section~\ref{sec:calibration}, we apply it to four additional workloads spanning both local and global dependency classes.

\paragraph{Bounded-Degree Dependency}
Distributed SpMV computes $y \leftarrow Ax$ for a sparse matrix $A$ arising from a five-point Laplacian discretization on a 2D grid, distributed by rows across ranks. Each rank requires boundary values only from its immediate neighbors, inducing nearest-neighbor halo exchanges identical in structure to the stencil workload. Conv2D applies a fixed-radius convolution over a spatial grid, similarly requiring only nearest-neighbor halo exchanges~\cite{lavin2016fast,chetlur2014cudnn}. In both cases, each rank communicates with a fixed number of neighbors regardless of total rank count, so the number of dependency edges per phase grows linearly in $P$. We therefore model their scheduling overhead using Eq.\ref{eq:stencil_overhead}.

\begin{figure}[t]
    \centering
    \includegraphics[width=\linewidth]{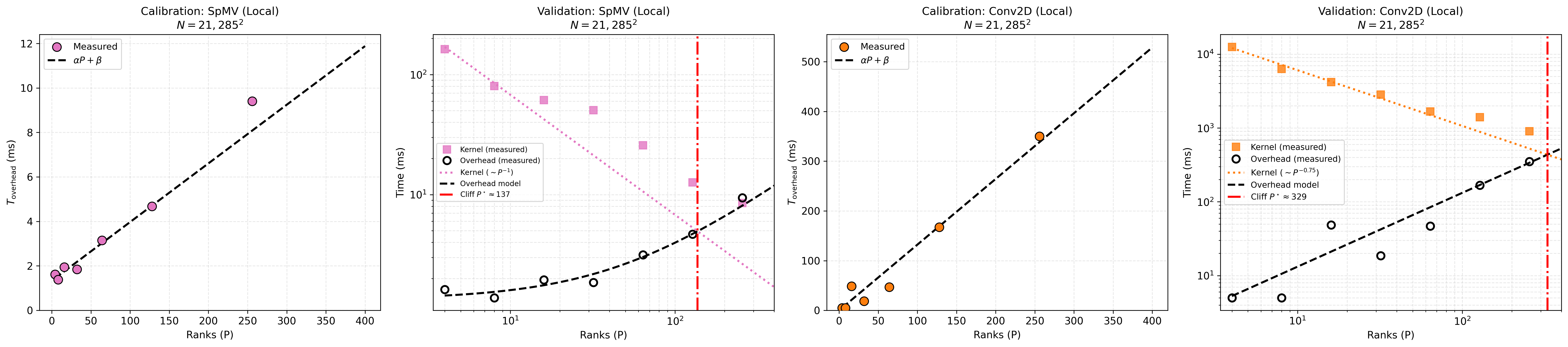}
    \includegraphics[width=\linewidth]{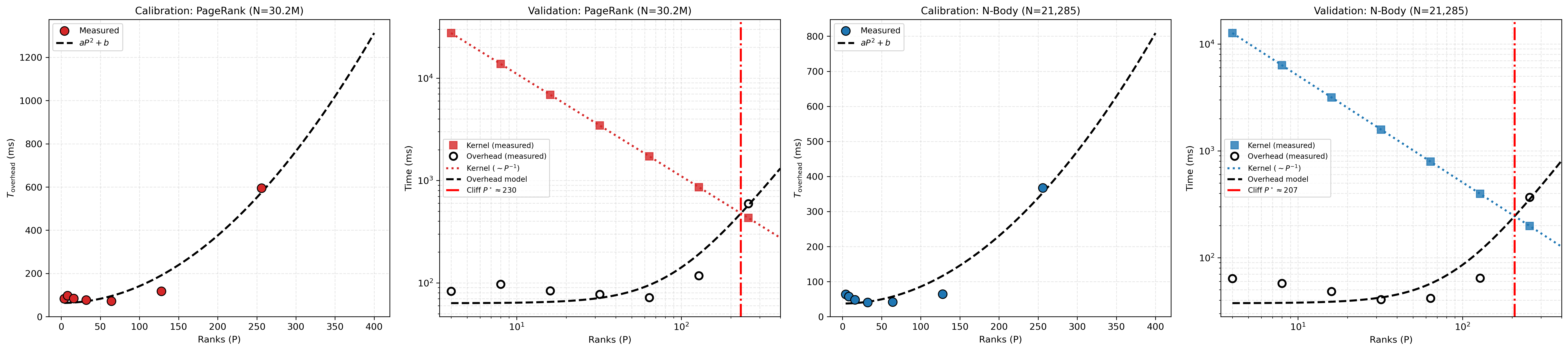}
    \caption{Calibration and validation of the scheduling-overhead model for four additional workloads. Top row: SpMV uses a five-point Laplacian on a 2D grid with 1D row decomposition and nearest-neighbor halo exchange; Conv2D uses 1D row decomposition with a $3\times3$ kernel, $C_{\mathrm{in}}=3$ input channels, and $C_{\mathrm{out}}=16$ output channels. Bottom row: PageRank and N-Body. All workloads are configured for comparable total computational work (${\sim}453$M operations per iteration): SpMV and Conv2D operate on a $21{,}285^2$ grid, N-Body computes $O(N^2)$ pairwise interactions among $N=21{,}285$ particles, and PageRank processes $N=30.2$M vertices with average degree 15 (${\sim}453$M edges).}
    \label{fig:additional_prediction}
\end{figure}

Figure~\ref{fig:additional_prediction} (top row) evaluates the proposed linear overhead model for SpMV and Conv2D. For both workloads, measured scheduling overhead increases approximately linearly with rank count, and the fitted $\alpha P + \beta$ model closely matches the measurements. For SpMV, the predicted crossover occurs at $P^\star \approx 137$. For Conv2D, the crossover is predicted at a larger scale ($P^\star \approx 329$), reflecting the higher per-rank computational cost of the convolution kernel, which delays the point at which overhead overtakes useful computation.

\paragraph{Dense Dependency}
PageRank and all-pairs N-Body simulation both exhibit dense inter-rank dependency structure. In our distributed PageRank implementation~\cite{page1999pagerank,malewicz2010pregel}, each iteration performs a sparse matrix--vector product followed by a global normalization step. The graph's adjacency structure induces cross-rank data dependencies, in which each rank must exchange updated vertex values with all ranks that own destination vertices. For graphs whose adjacency structure spans most partitions, this produces $O(P)$ dependencies per rank and $O(P^2)$ total dependency edges per iteration. All-pairs N-Body simulation~\cite{nyland2007fast,aarseth2003gravitational} computes gravitational forces between $N$ particles. Under domain decomposition, each rank owns $N/P$ particles but requires the positions of \emph{all} $N$ particles to compute local forces. This all-to-all data dependency creates $O(P)$ edges per rank and $O(P^2)$ total edges per phase. We therefore model scheduling overhead for both workloads using Eq.\ref{eq:fft_overhead}.

To ensure comparable kernel execution times across workloads, we select input sizes that produce similar total computational work: N-Body with $N = 21{,}285$ particles performs $O(N^2) \approx 453$M pairwise force computations, while PageRank with $N = 30.2$M vertices and average degree 15 traverses ${\sim}453$M edges per iteration. Figure~\ref{fig:additional_prediction} (bottom row) presents calibration and validation results for the largest problem sizes of PageRank and N-Body. In both cases, measured scheduling overhead follows the quadratic model with high fidelity, confirming that the $\alpha P^2 + \beta$ functional form predicted by dense dependency structure accurately captures overhead growth across algorithmically distinct workloads.

The validation panels overlay the calibrated overhead model against
measured kernel execution time under strong scaling. For PageRank, the predicted crossover occurs at $P^\star \approx 230$, while for N-Body the crossover occurs at $P^\star \approx 207$.

\subsection{Dynamic--Static Crossover Under Strong Scaling}
\label{sec:dynamic_static}

\begin{figure}[t]
    \centering
    \includegraphics[width=\linewidth]{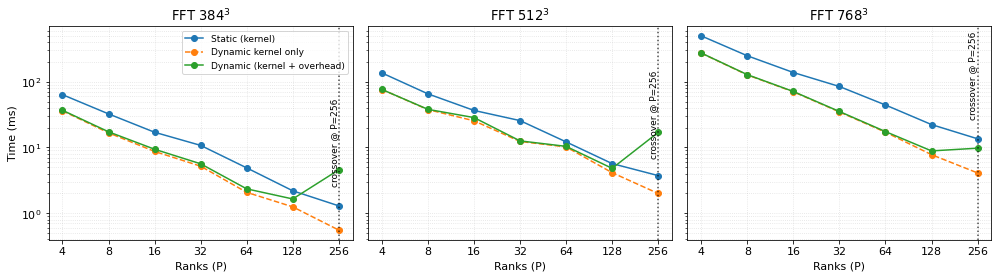}
    \caption{Static versus dynamic execution for FFT under strong scaling.
    Each plot shows the static kernel time, the dynamically scheduled kernel time, and the total dynamic execution time (kernel plus scheduling overhead).}
    \label{fig:dynamic_static_crossover}
\end{figure}

To connect the granularity-based characterization to practical scheduling decisions, Figure~\ref{fig:dynamic_static_crossover} directly compares static execution against dynamically scheduled execution for FFT across multiple input sizes. The dynamically scheduled kernel consistently outperforms the static kernel due to improved load balance. However, at sufficiently large scales, scheduling overhead grows rapidly and causes total dynamic execution time to exceed static execution time.

We observe a clear dynamic--static crossover for all FFT problem sizes, occurring at high rank counts, which occurs when the load-balance benefit of dynamic execution can no longer amortize scheduling overhead, $T_{\text{static}} = T_{\text{kernel,dyn}} + T_{\text{overhead}}$.
\paragraph{Static baseline.}
All measurements in this work isolate a single computation stage. We therefore compare dynamic execution against a matching static compute-only baseline. For one FFT computation stage, the static execution time under ideal load balance is modeled as:
\begin{equation}
\widehat{T}_{\text{static}}(P)
=
\frac{N^2}{P}\; c\; N \log N ,
\label{eq:static_fft}
\end{equation}
where $N$ is the global problem dimension, $P$ is the number of ranks, and $c$ is an implementation- and architecture-dependent constant capturing the cost of a single length-$N$ one-dimensional FFT~\cite{czechowski2012fft}.

\paragraph{Dynamic execution.}
Under dynamic scheduling, the predicted execution time for the same computation stage is:
\begin{equation}
\widehat{T}_{\text{dyn}}(P) = \widehat{T}_{\text{kernel}}(P) + \widehat{T}_{\text{overhead}}(P),
\label{eq:dynamic_fft}
\end{equation}
where \(\widehat{T}_{\text{kernel}}(P)\) models compute time and \(\widehat{T}_{\text{overhead}}(P)\) captures scheduler overhead. As shown earlier, FFT overhead grows superlinearly with rank count due to global dependency.

\paragraph{Runtime decision rule.}
Combining the static compute-only model (Eq.~\ref{eq:static_fft}) with the dynamic execution model (Eq.~\ref{eq:dynamic_fft}) yields a simple crossover-based scheduling criterion:
\begin{equation}
\text{use dynamic scheduling if }
\widehat{T}_{\text{kernel}}(P)
+
\widehat{T}_{\text{overhead}}(P)
<
\widehat{T}_{\text{static}}(P),
\label{eq:decision_rule}
\end{equation}
and otherwise prefer static execution.

Because $\widehat{T}_{\text{static}}(P)$ and $\widehat{T}_{\text{kernel}}(P)$ both scale approximately as $P^{-1}$, while $\widehat{T}_{\text{overhead}}(P)$ grows with coordination cost, this inequality predicts a finite crossover point $P^\star$ beyond which scheduling overhead can no longer be amortized.
Crucially, all terms in Eq.~\ref{eq:decision_rule} can be estimated at runtime using the scheduler’s profiling information and the calibrated overhead model, enabling informed scheduling decisions without requiring offline strong-scaling measurements.

\section{Summary}
\label{sec:summary}

This work shows that scheduling overhead in task-based runtimes is governed primarily by dependency structure rather than task count or problem size alone. By relating overhead growth directly to task-graph dependencies, we explain why some workloads exhibit abrupt strong-scaling breakdowns while others degrade gradually or remain stable. The proposed models are intentionally lightweight and are not meant to predict exact execution times, but to identify when scheduling overhead can no longer be amortized. The close alignment between model-predicted crossover points and measured performance degradation suggests that dependency-aware modeling is sufficient for reasoning about scheduling effectiveness without exhaustive scaling studies. This structural characterization enables principled decisions between dynamic and static execution based on task-graph dependency properties, without requiring exhaustive strong-scaling experimentation. Integrating dependency-aware overhead prediction directly into runtime schedulers is a promising direction for making task-based parallelism more robust at scale. Such integration would enable automatic transitions between dynamic and static execution as $G$ approaches the marginal regime.
%
%
%
%

\end{document}